\documentclass{article}
\usepackage{color}
\usepackage{graphics}
\usepackage{ascmac}
\usepackage{amsmath}
\usepackage{amssymb}
\usepackage{amsfonts}
\usepackage{hyperref}
\usepackage{tcolorbox}
\usepackage{multirow}
\usepackage{arydshln}
\usepackage{algorithm}
\usepackage{algorithmic}
\tcbuselibrary{breakable}
\usepackage{array}
\usepackage{longtable}
\usepackage{colortab}
\usepackage{colortbl}
\usepackage{authblk}
\usepackage{here}
\newcommand{\ket}[1]{|{#1}\rangle}
\newcommand{\kb}[1]{|{#1}\rangle\!\langle{#1}|}
\newcommand{\bk}[1]{\langle{#1}|{#1}\rangle}
\newcommand{\bra}[1]{\langle{#1}|}
\newcommand{\bkt}[2]{\langle{#1}|{#2}\rangle}
\newcommand{\kbt}[2]{|{#1}\rangle\!\langle{#2}|}
\newcommand{\ave}[1]{\langle{#1}\rangle}
\makeatother
\makeatletter

\begin{document}
\title{Post-Hartree-Fock method in Quantum Chemistry for Quantum Computer}
\author[1,2]{Yutaka Shikano}
\author[1,3]{Hiroshi C. Watanabe}
\author[4,1]{Ken M. Nakanishi}
\author[5,1]{Yu-ya Ohnishi}
\affil[1]{Quantum Computing Center, Keio University, 3-14-1 Hiyoshi, Kohoku, Yokohama, 223-8522, Japan}
\affil[2]{Institute for Quantum Studies, Chapman University, 1 University Dr., Orange, CA 92866, USA}
\affil[3]{JST PRESTO, 4-1-8 Honcho, Kawaguchi, Saitama 332-0012, Japan}
\affil[4]{Department of Physics, Graduate School of Science, The University of Tokyo, 7-3-1 Hongo, Bunkyo-ku, Tokyo 113-0033, Japan}
\affil[5]{Materials Informatics Initiative, RD Technology \& Digital Transformation Center, JSR Corporation, 100 Kawajiri-cho, Yokkaichi, Mie, 510-8552, Japan}
\date{} 
\maketitle
\begin{abstract}
    Quantum computational chemistry is a potential application of quantum computers that is expected to effectively solve several quantum-chemistry problems, particularly the electronic structure problem. Quantum computational chemistry can be compared to the conventional computational devices. This review comprehensively investigates the applications and overview of quantum computational chemistry, including a review of the Hartree-Fock method for quantum information scientists. Quantum algorithms, quantum phase estimation, and variational quantum eigensolver, have been applied to the post-Hartree-Fock method.
\end{abstract}

\section{Introduction} \label{sec:intro} 
Computer simulation initially applied in meteorology and nuclear physics is an alternative tool that can be used in theoretical modelling to compare experimental data. This powerful tool is essential for achieving scientific and engineering accomplishments along with the rapid development of computational devices. According to Ref.~\cite{Winsberg}, the following perspective of computer simulation was stated:
\begin{quote}
    Successful simulation studies do more than compute numbers. Instead, they utilize a variety of techniques to draw inferences from these numbers. The simulations make innovative use of the calculational techniques that can only be supported extra-mathematically and extra-theoretically. As such, unlike simple computations that can be conducted on a computer, the results of the simulations are not automatically reliable. Considerable effort and expertise are required to decide which simulation results are reliable.
\end{quote}
As an emerging technology for next-generation computing platforms, the simulations performed by quantum computers are expected to be used for scientific research and industrial applications~\cite{Alexeev}. In the quantum computing era, this perspective should remain unchanged. Although quantum computers cannot be regarded as autonomous research tools, they can be used an auxiliary device for achieving better understanding and more applications.

Walter Heinrich Heitler and Fritz Wolfgang London initiated the quantum mechanical calculation of bonding properties of the hydrogen molecule ${\rm H_2}$ in 1927~\cite{Heitler}. Thereafter, Linus Carl Pauling, who got the Nobel prize in chemistry in 1954, developed the fundamental concept of the chemical bonds by solving the many-body Schr\"{o}dinger equation~\cite{Pauling}. However, this is not analytically solved in general. To understand the fundamental properties of molecules and chemical reactions, approximate computational methods to solve the many-body Schr\"{o}dinger equation are developed along with the understanding of the properties of the many-body quantum system. This academic field for molecules is termed \textit{computational chemistry} or \textit{quantum chemistry}. 

Although several computational techniques and hardware improvements have been evolving, it remains challenging to numerically solve the many-body quantum system~\cite{Gordon}. On the other hand, the conceptual idea of quantum simulation~\cite{Feynman} and quantum computer~\cite{Deutsch} has inspired new computational algorithms. In 2005, a seminal paper proposed a quantum computing algorithm for computational chemistry, which resulted in the emergence of the field of quantum computational chemistry~\cite{Alan}. Quantum computational algorithms and techniques for quantum chemistry applications are under development. Furthermore, the applications of cloud computing in the quantum era or for quantum computing at the noisy intermediate-scale quantum (NISQ) era~\cite{Preskill} in real devices are being studied as seen in \cite[Table2]{qc_review}. Comprehensive review papers have been reported in previous studies~\cite{qc_review,qc_review2,qc_review3}. Therefore, the aim of this review is to explain the technical assessments of quantum computational chemistry, particularly, the computational accuracy and approximations for quantum information scientists reviewed in Ref.~\cite{Szabo}. Quantum computational methods for solving quantum many-body system are relevant in nuclear physics~\cite{Dumitrescu,Klco} and statistical physics~\cite{Smith}.

\begin{table}[t]
\centering
\begin{tabular}{|r|l|l|l|}
\hline
Sec. \ref{sec:problem}            & \multicolumn{3}{l|}{Problem Setting (Quantum many-body Hamiltonian)}                                                             \\ \hline
\multirow{2}{*}{$\downarrow$} & \multicolumn{1}{r|}{Sec. \ref{sec:appli}}    & \multicolumn{2}{l|}{Applications}                               \\ \cline{2-4} 
                  & \multicolumn{1}{r|}{Sec. \ref{sec:valid}}    & \multicolumn{2}{l|}{Validations}                                \\ \hline
Sec. \ref{sec:HF}            & \multicolumn{3}{l|}{Hartree-Fock (HF) method}                                                    \\ \hline
Sec. \ref{sec:post}            & \multicolumn{3}{l|}{Post HF methods}                                                             \\ \hline
Sec. \ref{sec:fullci}          & Full Configuration Interaction & \multicolumn{1}{r|}{Sec. \ref{sec:cc}} & Coupled Cluster method          \\ \hline
Sec. \ref{sec:qubit}            & \multicolumn{3}{l|}{Qubit mapping}                                                               \\ \hline \hline
Sec. \ref{sec:quantum}            & \multicolumn{3}{l|}{Quantum algorithms for post HF methods}                                      \\ \hline
Sec. \ref{sec:qpe}          & Quantum Phase Estimation       & \multicolumn{1}{r|}{Sec. \ref{sec:vqe}} & Variational Quantum Eigensolver \\ \hline \hline
Sec. \ref{sec:conclusion}            & \multicolumn{3}{l|}{Conclusion}                                                                  \\ \hline
\end{tabular}
\caption{Paper organization.}
\label{table:paper}
\end{table}
The rest of this paper is organized as seen in Table~\ref{table:paper}. Throughout this paper, the SI unit is used, unless otherwise stated. The constants $\epsilon_0$, $\hbar$, and $q_e$ are the permittivity of free space, the reduced Planck's constant, and the elementary charge, respectively.

\section{Problem Setting of Quantum Chemistry Calculation} \label{sec:problem}
The eigenvalue and eigenstate problem can be solved using the non-relativistic many-body Schr\"{o}dinger equation: 
\begin{equation}
    \label{origin}
    H \ket{\Psi} = E \ket{\Psi}.
\end{equation}
It is noted that the relativistic effect should be considered, especially for the heavier elements~\cite{Pyykko}. This can be taken as the first approximation of many-body quantum system to understand the chemical properties. This treatment is dealt as perturbations, or small corrections, to the non-relativistic theory of chemistry as seen the details in Ref.~\cite{Dyall}.
In quantum mechanics, the non-relativistic Hamiltonian is given by: 
\begin{equation}
    H = T + V, 
\end{equation}
where, the kinetic term is 
\begin{equation}
    T = - \frac{\hbar}{2 m_e} \sum_{i=1}^{N_e} \nabla^2_{i} - \sum_{I=1}^{N_{nucl}} \frac{\hbar}{2 m_{nucl, I}} \nabla^2_{I} \equiv T_e + T_{nucl}
\end{equation}
with $N_e$, $N_{nucl}$, $m_e$, and $m_{nucl}$ being the number of electrons, the number of nuclei, the mass of an electron, and the nuclear mass, respectively. Here, the electron and $I$-th nuclear masses are denoted as $m_e$ and $m_{nucl,I}$, respectively. The potential term, which included the electron-electron, nuclei-nuclei, and electron-nuclei interactions, is given by 
\begin{equation}
    V =  \frac{1}{4 \pi \epsilon_0} \sum_{i<j}^{N_e} \frac{q_e^2}{|\vec{x}_i - \vec{x}_j|} 
    - \frac{1}{4 \pi \epsilon_0} \sum_{I=1}^{N_{nucl}} \sum_{j=1}^{N_e} \frac{Z_I q_e^2}{|\vec{X}_I - \vec{x}_j|} 
    + \frac{1}{4 \pi \epsilon_0} \sum_{I<J}^{N_{nucl}} \frac{Z_I Z_J q_e^2}{|\vec{X}_I - \vec{X}_J|},
\end{equation}
where $\vec{x}_i$ and $\vec{X}_I$ are the electron and nuclei coordinates, respectively; and $Z_I q_e$ is the charge of nuclei. As previously mentioned, this eigenvalue and eigenstate problem cannot be analytically solved. Large computational resources require directly solving the aforementioned challenges, even if it involves using numerical methods. Subsequently, several approximations are utilized.

As a first approximation, we consider the scenario that the motion of atomic nuclei and electrons in a molecule can be treated as separate entities. This means that the entire wavefunction of a many-body system 
\begin{equation*}
    \ket{\Psi (\{ \vec{x}_i \} \equiv \{ \vec{x}_1, \cdots, \vec{x}_{N_e} \}, \{ \vec{X}_I \} \equiv \{ \vec{X}_1, \cdots, \vec{X}_{N_{nucl}} \})}
\end{equation*} 
can be approximately decomposed to
\begin{equation}
    \ket{\Psi (\{ \vec{x}_i \}, \{ \vec{X}_I \})} \approx \ket{\psi_{e} (\{ \vec{x}_i \}; \{ \vec{X}_I \})} \ket{\psi_{nucl} (\{ \vec{X}_I \})}.
\end{equation}
This is often referred to as the clamped-nuclei approximation. Under this assumption, the original eigenvalue and eigenstate problem is divided to the two eigenvalue and eigenstate problems; 
\begin{equation}
    \tilde{H}_{e} \ket{\psi_{e} (\{ \vec{x}_i \}; \{ \vec{X}_I \})} = V_{e} (\{\vec{X}_I \}) \ket{\psi_{e} (\{ \vec{x}_i \}; \{ \vec{X}_I \})}, \label{elech}
\end{equation}
where the electronic Hamiltonian ($\tilde{H}_e$) neglects the nuclear kinetic term ($T_n$). Under the given nuclear coordinate ($\{ \vec{X}_I \}$), Eq. (\ref{elech}) is solved. Then, the solved eigenvalue ($V_{e} (\{\vec{X}_I \})$) used for varying the nuclear coordinate ($\{ \vec{X}_I \}$) is termed the interatomic potential or the potential energy surface. This is often referred to as the adiabatic approximation. Thereafter, we solve the second problem for the nuclear motion as
\begin{equation}
    (T_n + V_{e} (\{\vec{X}_I \})) \ket{\psi_{nucl} (\{ \vec{X}_I \})} = E \ket{\psi_{nucl} (\{ \vec{X}_I \})}. \label{nuclh}
\end{equation}
Due to the Eckart condition~\cite{Eckart}, the vibrational, translational, and rotational motions of the molecule can be separated. The solved eigenvalue ($E$) represents the total energy of the molecule. The entire procedure is termed the Born-Oppenheimer (BO) approximation. This approximation is justified when the energy gap between the ground and excited electronic states is larger than the energy scale of the nuclear motion. Therefore, this approximation loses validity in the instances of zero band gap, vibronic coupling in electronic transitions (Herzberg-Teller effect), ground state degeneracies removed by lowering the symmetries (Jahn-Teller effect), and the interaction of electronic and vibrational angular momenta (Renner-Teller effect). For example, metals, graphene, and topological materials exhibit a zero band gap. The octahedral complexes of transition metals such as six-coordinate copper (II) complexes usually correspond to the Jahn-Teller effect. There are several treatments on non-BO approximation in quantum chemistry ~\cite{Bubin,Takatsuka}. In quantum computation, this generalization was considered in a previous study ~\cite{Nakai}.

Without loss of generality, we consider the following electronic Hamiltonian for the fixed nuclear coordinate ($\{ \vec{X}_I \}$),
\begin{equation}
    \label{eHam}
    H_e = - \frac{\hbar}{2 m_e} \sum_{i=1}^{N_e} \nabla^2_{i} + \frac{1}{4 \pi \epsilon_0} \sum_{i<j}^{N_e} \frac{q_e^2}{|\vec{x}_i - \vec{x}_j|} 
    - \frac{1}{4 \pi \epsilon_0} \sum_{I=1}^{N_{nucl}} \sum_{j=1}^{N_e} \frac{Z_I q_e^2}{|\vec{X}_I - \vec{x}_j|}.
\end{equation}
Therefore, we focus on solving the eigenvalue ($E_n (\{ \vec{X}_I \})$) and its corresponding eigenstate ($\ket{\psi_{n,e} (\{ \vec{x}_i \}; \{ \vec{X}_I \})}$) with the discrete index, $n = 0, 1, \cdots$, as
\begin{equation}
    \label{problem}
    H_e \ket{\psi_{n,e} (\{ \vec{x}_i \}; \{ \vec{X}_I \})} = E_n (\{ \vec{X}_I \}) \ket{\psi_{n,e} (\{ \vec{x}_i \}; \{ \vec{X}_I \})}.
\end{equation}
This is often referred to as the electronic structure calculation. Our primary objective is to obtain the eigenvalue as well as its corresponding eigenstate. The case of $n=0$ corresponds to the ground state of the molecule. For simplicity, we focus on the ground state throughout this paper.

In computational chemistry, the intermolecular distance is often normalized by the Bohr radius, $a_0 \equiv 4 \pi \epsilon_0 \hbar^2 / (m_e q_e^2) = 0.529$ \AA, which is exactly equal to the most probable distance between the nucleus and the electron in a hydrogen atom in its ground state. The obtained eigenvalue, the electronic energy, uses the Hartree equation, $E_h \equiv \hbar^2 / (m_e a_0^2) = 27.2 ~{\rm eV} = 4.36 \times 10^{-18}~{\rm J} = 2625~{\rm kJ / mol}$, which is equivalent to the electric potential energy of the hydrogen atom in its ground state and, by the virial theorem, approximately twice its ionization energy. 

\section{Applications of Electronic Structure Calculation} \label{sec:appli}
\begin{figure}[t]
    \centering
    \includegraphics{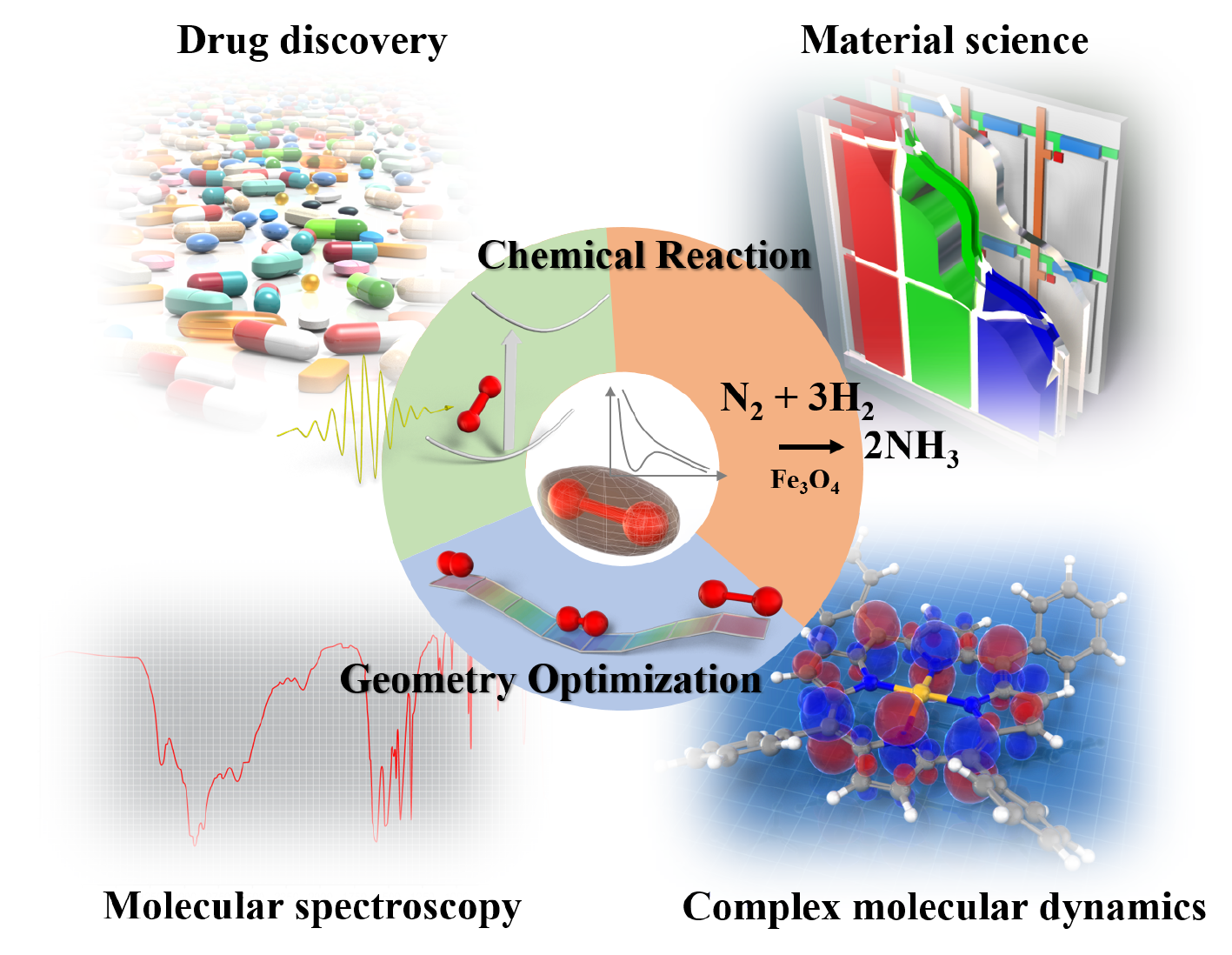}
    \caption{Various potential applications of computational chemistry: drug and material designs, understanding molecular functionality, and earth and atmosphere science.}
    \label{fig:qc2}
\end{figure}

At the end of his Nobel Lecture in 1966~\cite{Mulliken}, Robert S. Mulliken stated: 
\begin{quote}
    In conclusion, I would like to emphasize my belief that the era of computing chemists, when hundreds if not thousands of chemists will go to the computing machine instead of the laboratory, for increasingly many facets of chemical information, is already at hand. There is only one obstacle, namely, that someone must pay for the computing time.
\end{quote}
\begin{figure}[t]
    \centering
    \includegraphics{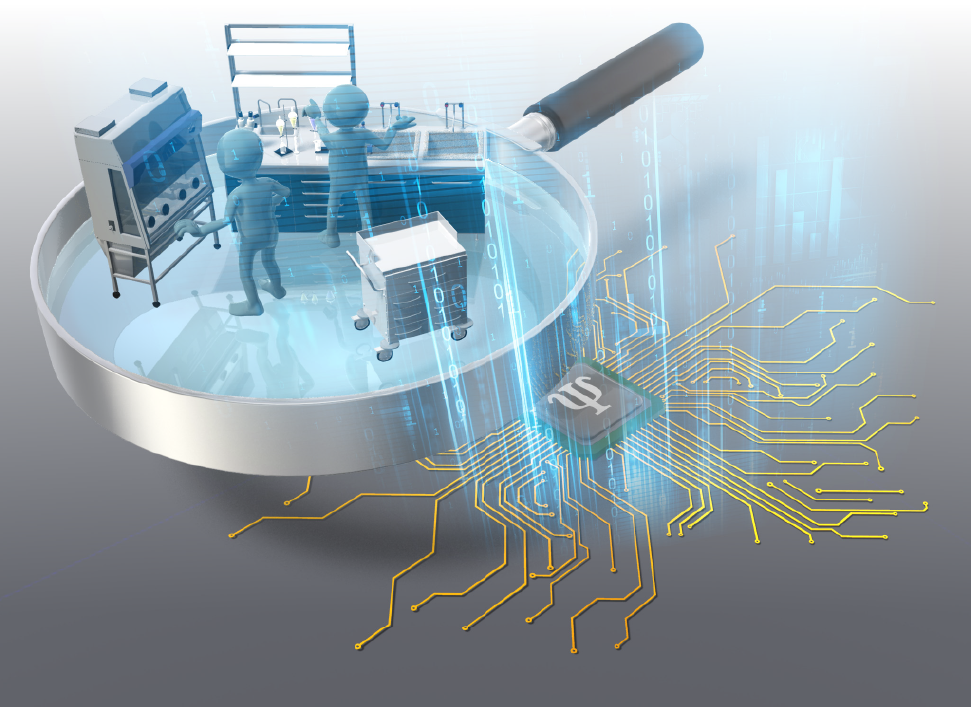}
    \caption{The dream of computational chemistry: substituting a (quantum) computing chip for a chemical laboratory.}
    \label{fig:qc}
\end{figure}
The ultimate goal of computational chemistry is to simulate chemical phenomena in test tubes in a laboratory by numerically solving the many-body Schr\"{o}dinger equation (\ref{origin}) in computational chips, as depicted in Fig.~\ref{fig:qc}. In the following subsections, the exact or approximate solution to Eq.~(\ref{problem}) is applied to the static properties of molecules and the chemical reactions including chemical dynamics, as depicted in Fig.~\ref{fig:qc2}.

A chemical reaction is a process that results in the chemical transformation of one set of chemical substances into another. One of the ultimate goals in chemistry is the comprehension of various chemical dynamics. The transition state theory explains the reaction rates of elementary chemical reactions due to a structural change. The potential energy surface between reactants and products requires locating the reaction pathway~\cite{TES}. The maximum-energy state of the reaction pathway is called a transition state. To reduce the computational cost of the entire potential energy surface, computational algorithms have been recently developed to find the reaction pathway~\cite{Maeda}. Instead of the transition state theory, electron-transfer reactions such as redox reactions are described by the Marcus theory~\cite{Marcus}. The comprehension of more complicated chemical reactions in a condensed phase necessitates the establishment of a post-Marcus theory such as the quantum coherence enhanced theory~\cite{quantumcoherence}.

Geometry optimization to capture a molecular structure is one of the most important techniques in the field of quantum chemical calculation when one tries to interpret chemical phenomena, as observed in the review article~\cite{Schlegel}. We obtain the stable ground-state energy of Eq. (\ref{problem}) with variables of the nuclei ($\{ \vec{X}_I \}$) to optimize the potential energy surface (PES). To solve the stable ground-state energy, the first derivative of the PES with respect to variables of the nuclei ($\{ \vec{X}_I \}$), which corresponds to the negative of the force, is calculated. The zero of this first derivative includes the maxima; minima; first order saddle points, which are called transition states; and higher order saddle points of the PES. Thus, its second derivatives with respect to the variables of the nuclei ($\{ \vec{X}_I \}$) are obtained. The derivatives with respect to $N_{nucl}$ nuclear positions are calculated, and the $N_{nucl}$-th order square matrix is obtained and is called a Hessian matrix. These eigenvalues are proportional to the square of the vibrational spectra, and its corresponding eigenvectors are the normal modes of molecular vibration. In most cases, the energy difference between these states semi-quantitatively agrees with experimental observation when the highly sophisticated computational methods have large basis sets. Even less accurate computational methods like density functional theory (DFT)~\cite{DFT} can present qualitatively comparable energy differences compared to experimental methods, once the electronic structure of the system is not extremely complex. 
\section{Validation of Computational Techniques} \label{sec:valid}
\subsection{Comparison to experimental results}
The energy eigenvalue, that is, the exact or approximate solution to Eq. (\ref{problem}), itself is not directly observed, while the energy difference is observable. In theory, the expected value of the physical observable $\hat{O}$ can be computed as 
\begin{equation}
    \ave{\hat{O}} = \bra{\psi_{n,e} (\{ \vec{x}_i \}; \{ \vec{X}_I \})} \hat{O} \ket{\psi_{n,e} (\{ \vec{x}_i \}; \{ \vec{X}_I \})}.
\end{equation}
The electronic dipole moment of molecule $\mu_e$ is defined as 
\begin{equation}
    \mu_e := e \ave{\hat{x}} = e \bra{\psi_{n,e} (\{ \vec{x}_i \}; \{ \vec{X}_I \})} \hat{x} \ket{\psi_{n,e} (\{ \vec{x}_i \}; \{ \vec{X}_I \})},
\end{equation}
where $\hat{x}$ is the position operator for the electrons, which represents the chemical polarity. Furthermore, the laser spectroscopy is an important technique for observing the chemical phenomena. The transition dipole moment of molecules $\mu_T$ in the quantum state $\ket{\psi_{n,e} (\{ \vec{x}_i \}; \{ \vec{X}_I \})}$ to $\ket{\psi_{n^\prime,e} (\{ \vec{x}_{i^\prime} \}; \{ \vec{X}_{I^\prime} \})}$ is defined as
\begin{equation}
    \mu_T := e \bra{\psi_{n^\prime,e} (\{ \vec{x}_{i^\prime} \}; \{ \vec{X}_{I^\prime} \})} \hat{x} \ket{\psi_{n,e} (\{ \vec{x}_i \}; \{ \vec{X}_I \})}.
\end{equation}
This quantity is directly verified by absorption or transmission spectroscopy to obtain the energy spectrum of the Hamiltonian (\ref{problem}). Furthermore, the vibrational motion of the nuclei, which is termed a molecular vibration, induces changes in the electronic and transition dipole moments such as $d \mu_e / d \vec{X}_R$ and $d \mu_T / d \vec{X}_R $ with the fundamental vibrational coordinate, $\vec{X}_R$. This is verified by vibrational spectroscopy such as Raman spectroscopy. Therefore, the energy eigenstate of the Hamiltonian (\ref{problem}) provides the transition properties of molecules as well as the spectroscopic information to be compared with the experimental values.
\subsection{Precision criteria of electronic structure calculation} \label{sec:accuracy}
According to the Nobel Lecture by John Pople in 1998~\cite{Pople},
\begin{quote}
    A target accuracy must be selected. A model is not likely to be of much value unless it is able to provide clear distinction between possible different modes of molecular behavior. As the model becomes quantitative, the target should be to reproduce and predict data within the experimental accuracy. For energies, such as heats of formation or ionization potentials, a global accuracy of 1 kcal/mole would be appropriate.
\end{quote}
The target accuracy of computational chemistry strongly depends on the demand of the application. Conventionally, when considering a gas-phase reaction at room temperature, the molecules are approximately equilibrated. The thermal energy of room temperature is ~0.6 kcal/mol. Therefore, 1 kcal/mol = 1.6 mhartree, which has been termed as the \textit{chemical accuracy}, is often set as the target accuracy of computational accuracy.

\section{Hartree-Fock method}
\label{sec:HF}
This method is essentially the mean-field theory for electrons. The method can be used to solve the optimized single electron wavefunction (i.e., the molecular orbital), under the condition that dynamics of this single electron are susceptible to the nucleus and the effective potential formed by the surrounding electrons. Therefore, this method can be regarded as the approximation that disregards the electron correlation when solving the electronic structure problem, Eq. (\ref{problem}).

The Hartree-Fock approximation for the $N_e$ electron system is the anti-symmetric quantum state $\ket{\psi_{e} (\{ \vec{x}_i \}; \{ \vec{x}_I \})}$, which uses a single Slater determinant, 
\begin{align}
    \ket{\psi_{e} (\{ \vec{x}_i \}; \{ \vec{x}_I \})} \approx \ket{\psi_{HF} (\{ \vec{x}_i \})} & \equiv \frac{1}{\sqrt{N_e!}} \left|
	\begin{array}{cccc}
	\varphi_{1}(\vec{x}_1) & \varphi_{1}(\vec{x}_2) & \ldots & \varphi_{1}(\vec{x}_{N_e}) \\
    \varphi_{2}(\vec{x}_1) & \varphi_{2}(\vec{x}_2) & \ldots & \varphi_{2}(\vec{x}_{N_e}) \\
    \vdots & \vdots & \ddots & \vdots \\
    \varphi_{N_e}(\vec{x}_1) & \varphi_{N_e}(\vec{x}_2) & \ldots & \varphi_{N_e}(\vec{x}_{N_e}) \\
    \end{array} \right| \\ 
    & \equiv \ket{\varphi_{1} (\{ \vec{x}_i \})} \ket{\varphi_{2} (\{ \vec{x}_i \})} \cdots \ket{\varphi_{N_e} (\{ \vec{x}_i \})}. \label{slater_rep}
\end{align}
This is because the electron is a fermion. This is also termed a single-determinant approximation. In addition, $\{ \ket{\varphi_{i} (\{ \vec{x}_j \})} \}$ is called a molecular orbital (MO). Under this assumption, given the trial Hartree-Fock (HF) state $\ket{\psi_{HF} (\{ \vec{x}_i \})}$ with $\{ \varphi_i (\vec{x}) \}$, the energy functional, $E^{HF}_e [\{ \ket{\varphi_i (\{ \vec{x}_i \})} \}]$, is decomposed to 
\begin{equation}
    \label{ef}
    E^{HF}_e [\{ \ket{\varphi_i (\{ \vec{x}_i \})} \}] \equiv \bra{\psi_{HF} (\{ \vec{x}_i \})} H_e \ket{\psi_{HF} (\{ \vec{x}_i \})} = \sum_{i=1}^{N_e} h_{ii} + \frac{1}{2} \sum_{i,j=1}^{N_e} (J_{ij} - K_{ij}),
\end{equation}
where $h_{ii}$ is the single electron energy term, which is given by
\begin{equation}
    h_{ii} = \int d \vec{x} \varphi^\ast_{i} (\vec{x})
    \left( - \frac{\hbar}{2 m_e} \nabla^2  
    - \frac{1}{4 \pi \epsilon_0} \sum_{I=1}^{N_{nucl}} \frac{Z_I q_e^2}{|\vec{X}_I - \vec{x}|} \right)
    \varphi_{i} (\vec{x}),
\end{equation}
$J_{ij}$ is the Coulomb integral and $K_{ij}$ is the exchange integral given by
\begin{align}
    J_{ij} & = \iint d \vec{x} d \vec{\tilde{x}} \varphi^\ast_{i} (\vec{x}) \varphi^\ast_{j} (\vec{\tilde{x}})  \left( \frac{1}{4 \pi \epsilon_0} \frac{q_e^2}{|\vec{x} - \vec{\tilde{x}}|} \right) \varphi_{i} (\vec{x}) \varphi_{j} (\vec{\tilde{x}}) \\
    K_{ij} & = \iint d \vec{x} d \vec{\tilde{x}} \varphi^\ast_{i} (\vec{\tilde{x}}) \varphi^\ast_{j} (\vec{x})  \left( \frac{1}{4 \pi \epsilon_0} \frac{q_e^2}{|\vec{x} - \vec{\tilde{x}}|} \right) \varphi_{i} (\vec{x}) \varphi_{j} (\vec{\tilde{x}}),
\end{align}
respectively. When the variational principle is applied to the Lagrangian;
\begin{align}
    {\cal L} [\{ \ket{\varphi_i (\{ \vec{x}_i \})} \}] = \sum_{i=1}^{N_e} h_{ii} & + \frac{1}{2} \sum_{i,j=1}^{N_e} (J_{ij} - K_{ij}) \notag \\ & -  \sum_{i,j=1}^{N_e} \eta_{ij} \left( \bkt{\varphi_i (\{ \vec{x}_k \})}{\varphi_j (\{ \vec{x}_k \})} - \delta_{i,j} \right),
\end{align}
where the Lagrangian multiplier is represented by $\eta_{ij}$. Because $h_{ii}$ is the hermitian, we can derive the canonical HF equation,
\begin{equation}
    \label{HFeq}
    F \ket{\varphi_j (\{ \vec{x}_k \})} = \epsilon_j \ket{\varphi_j (\{ \vec{x}_k \})},
\end{equation}
where $\epsilon_i$ is the orbital energy and the Fock operator, $F$, is given by: 
\begin{align}
    F & = h + \sum_{i=1}^{N_e} \left( J_i - K_i \right), \label{canele} \\
    h & = - \frac{\hbar}{2 m_e} \nabla^2  
    - \frac{1}{4 \pi \epsilon_0} \sum_{I=1}^{N_{nucl}} \frac{Z_I q_e^2}{|\vec{X}_I - \vec{x}|}, \\
    J_i \varphi_j (\vec{x}) & = \int d \vec{\tilde{x}} \varphi^\ast_i (\vec{x}) \left( \frac{1}{4 \pi \epsilon_0} \frac{q_e^2}{|\vec{\tilde{x}} - \vec{x}|} \right) \varphi_i (\vec{\tilde{x}}) \varphi_j (\vec{x}), \label{coulomb} \\
    K_i \varphi_j (\vec{x}) & = \int d \vec{\tilde{x}} \varphi^\ast_i (\vec{\tilde{x}}) \left( \frac{1}{4 \pi \epsilon_0} \frac{q_e^2}{|\vec{\tilde{x}} - \vec{x}|} \right) \varphi_j (\vec{\tilde{x}}) \varphi_i (\vec{x}). \label{exhange}
\end{align}
This is the $N_e$ simultaneous nonlinear integral equation, which is termed the mean-field approximation. Even for the numerics, it is difficult to be solved. Next, the $N_e$ simultaneous nonlinear integral equation is mapped to the simultaneous algebraic equations by an additional approximation. 

\subsection{Basis sets approximation and Hartree-Fock-Roothaan equation}
The fundamental idea is that the MO $\{ \ket{\varphi_{i} (\{ \vec{x}_j \})} \}$ is approximately the linear combination of a given basis. This expression is given by 
\begin{equation}
    \label{linear}
    \ket{\varphi_{i} (\{ \vec{x}_j \})} \simeq \sum_{k=1}^M c_{ki} \ket{\chi_k},
\end{equation}
where $\{ \ket{\chi_k} \}_{k=1}^{M}$ is the basis set, and the coefficients $c_{ki}$ are unknown parameters. $M$ is the number of the basis set. It is noted that this linearly combined quantum state is not necessary to cover the entire Hilbert space of the single electron and is therefore an approximation. When the atomic orbitals (AO) are applied as the basis set, the result is the linear combination of atomic orbitals (LCAO) approximation. The atomic orbital is a mathematical function that describes the wave-like behavior of either one electron or a pair of electrons in an atom. This is based on the analytical solution of the eigenvalue and the eigenstate problem for the hydrogen atom. Therefore, the atomic orbital has the following three types; 
\begin{enumerate}
        \item The hydrogen-like atomic orbital,
    \item The Slater-type orbital (STO)---a form without radial nodes but decays from the nucleus like the hydrogen-like orbital,
    \item The Gaussian-type orbital (Gaussians)---no radial nodes and decays as $e^{- \alpha r^2}$ with the constant parameter $\alpha$ and the radial distance $r$. This is because the Coulomb and exchange integrals (\ref{coulomb}, \ref{exhange}) are quickly computed.
\end{enumerate}
Furthermore, the plane-wave basis sets are popular in calculations involving three-dimensional periodic boundary conditions. Under the polar coordinate system, the MO can be decomposed to 
\begin{equation}
    \chi_k (\vec{r}) = \bkt{\vec{r}}{\chi_k} =  
    \begin{cases}
        N_{\textrm{STO}} \cdot \exp (-\alpha_k r) Y_{l,m} (\theta, \phi), & (\textrm{Slater}) \\
        N_{\textrm{GTO}} \cdot \exp (-\alpha_k r^2) Y_{l,m} (\theta, \phi), & (\textrm{Gaussian}) \\
        N_{\textrm{PW}} \cdot \exp (- \vec{G}_k \cdot \vec{r}), & (\textrm{plane--wave})
    \end{cases}
\end{equation}
where $N_{\textrm{STO}}, N_{\textrm{GTO}}$, and $N_{\textrm{PW}}$ are the normalized constants; $Y_{l,m} (\theta, \phi)$ is the angular part of the wave function; $\alpha_k$ is the orbital constant; and $\vec{G}_k$ is the reciprocal lattice vector. Because the several basis sets are not easily computed while maintaining the computational accuracy, there are several types of basis sets.
\begin{tcolorbox}[title=Basis sets (Examples), breakable]
\begin{enumerate}
    \item {\it Minimal basis sets}: STO--$n$G
    \begin{itemize}
        \item $n$: the number of primitive Gaussian orbitals, which are fitted to a single Slater-type orbital (STO).
    \end{itemize}
    \begin{equation}
        \ket{\chi^{{\rm STO}}_k} = \sum^n_{m=1} \beta_{k, m} \ket{\chi^{{\rm GTO}}_m},
    \end{equation}
    where $\{ \beta_{k, m} \}$ is the fixed constant.
    \item {\it Pople's split-valence basis sets}: $X$--$YZ$G, $X$--$YZ$G*, or $X$--$YZ$+G
    \begin{itemize}
        \item $X$: the number of primitive Gaussians comprising each core atomic orbital basis function.
        \item $Y, Z$: the number of primitive Gaussian functions for the first and second valence STOs with the double zeta, repectively. The double-zeta representation is given by 
        \begin{equation}
            \ket{\chi_k} = \ket{\chi^{{\rm first \ STO}}_k} + d_k \ket{\chi^{{\rm second \ STO}}_k}
        \end{equation}
        with the fixed constants $\{ d_k \}$.
        \item *: with polarization functions on atoms in the second or later period.
        \item **: with the polarizing functions on hydrogen and atoms in the second or later period.
        \item +g: with diffuse functions on atoms in the second or later period.
        \item ++g: with diffuse functions on hydrogen and atoms in the second or later period.
    \end{itemize}
\end{enumerate}
\end{tcolorbox}
The choice of the basis sets determines the accuracy of the eigenvalue and its corresponding eigenstate, which will be discussed later. The details on the selection of the basis sets are provided in Refs. \cite{Davidson,Jensen,Helgaker}.

After selecting the basis sets, substituting Eq.~(\ref{linear}) into Eq. (\ref{HFeq}), and operating $\bra{\chi_j}$, we obtain the $M$ simultaneous algebraic equations as
\begin{align}
    \sum_{k=1}^M c_{ki} \bra{\chi_j} F \ket{\chi_i} & = \epsilon_i \sum_{k=1}^M c_{ki} \bkt{\chi_j}{\chi_i} \\
    \sum_{k=1}^M c_{ki} F_{ji} & = \epsilon_i \sum_{k=1}^M c_{ki} S_{ji}
\end{align}
with $F_{ji} \equiv \bra{\chi_j} F \ket{\chi_i}$ and $S_{ji} = \bkt{\chi_j}{\chi_i}$. This is termed the Hartree-Fock-Roothaan equation. 
For a non-trivial solution of the unknown parameters $c_{ki}$, 
\begin{equation}
    \label{HFR}
    \det ( F_{ji} - \epsilon_i S_{ji} ) = 0.
\end{equation}
The Hartree-Fock-Roothaan (HFR) equation is converted to the matrix equation: 
\begin{equation}
    \vec{F} \vec{c} = \vec{\epsilon} \vec{S} \vec{c},
    \label{HFRM}
\end{equation}
where $\vec{F} \equiv \{ F_{ji} \}$ is the Fock matrix, $\vec{S} \equiv \{ S_{ji} \}$ is the overlap matrix, and $\vec{c} \equiv \{ c_{ji} \}$ is the coefficient matrix. Here, $\vec{\epsilon}$ is the diagonal matrix of the orbital energies, $\epsilon_{i}$. This is solved using the following iterative process;
\begin{enumerate}
    \item Selecting the basis set. $\{ F_{ji} \}$ and $\{ S_{ji} \}$ are calculated.
    \item Initially guessing the parameters $\{ c_{ki} \}$.
    \item Solving Eq.~(\ref{HFR}) to obtain the estimated orbital energies $\{ \epsilon_i \}$.  \label{stepHR}
    \item Solving Eq.~(\ref{HFRM}) with the given parameters $\{ \epsilon_i \}$, updating the parameters $\{ c_{ki} \}$. 
    \item Repeat Step \ref{stepHR} until the parameter $\{ \epsilon_i \}$ converges.
\end{enumerate}
Therefore, this is often termed the self-consistent equation. The orbital energies, $\{ \epsilon_i \}$, and its corresponding approximated eigenstate are obtained since the parameters, $\{ c_{ki} \}$, are also solved. 

In summary, to solve quantum many-body problems with $N_e$ electrons, we make the following approximations; 
\begin{tcolorbox}
\begin{enumerate}
    \item Neglecting the relativistic effects, \label{A:RE}
    \item Born-Oppenheimer approximation, \label{A:BO}
    \item Hartree-Fock approximation, \label{A:HF}
    \item Mean-field approximation, \label{A:MF}
    \item Basis set approximation. \label{A:BS}
\end{enumerate}
\end{tcolorbox}
\subsection{Spin Coordinate}
The electron has the spin of $\frac{1}{2}$ as an intrinsic property. Therefore, the MO can be expressed as
    \begin{equation}
        \ket{\varphi_{i} (\{ \vec{x}_j \})} = \ket{\varphi_{i} (\{ \vec{r}_j \})} \ket{\alpha_i} 
    \end{equation}
    or 
    \begin{equation}
        \ket{\varphi_{i} (\{ \vec{x}_j \})} = \ket{\varphi_{i} (\{ \vec{r}_j \})} \ket{\beta_i}, 
    \end{equation}
where $\vec{r}_i$ is the electron coordinate and the spin variables, spin-up and spin-down, are denoted as $\alpha$ and $\beta$, respectively. When the spin coordinates are integrated, the canonical HF equation (\ref{HFeq}) can replace the electronic coordinates ($\{ \vec{x}_k \}$ to $\{ \vec{r}_k \}$) and the number of the electrons ($N_e$ to $N_e/2$) in Eq. (\ref{canele}). This treatment is called a restricted Hartree-Fock (RHF) method. This means that the exchange interactions among the spins are negligible. On the other hand, the spin-up and spin-down MOs are each independently computed. This is called an unrestricted Hartree-Fock (UHF) method. Importantly, we observed that a single Slater determinant of different orbitals for different spins is not a satisfactory eigenfunction of the total spin operator. This differentiation of the average value of the total spin operator is called a spin contamination.

Like noble gases, the MO is doubly occupied or empty, which is called a closed-shell configuration. The RHF method is applied. The other configurations are called open-shell configurations. The UHF method is applied. Otherwise, the restricted open-shell Hartree-Fock (ROHF) method, which assumes that the spin-up and spin-down MO energies are equal is applied. The details are provided in Ref.~\cite{Roothaan}.

\section{post-Hartree-Fock method} \label{sec:post}
In the previous section, several approximations of the HF method are discussed. The difference between the exact solution of Eq. (\ref{problem}) under the non-relativistic and BO assumptions and the HF solution arises from an electron correlation, which indicates the interactions among electrons. Therefore, the HF limit is the solution of Eq.~(\ref{HFeq}), which neglects the basis set approximation, and is always larger than the exact solution of Eq. (\ref{problem}). The energy difference is called a correlation energy. 

The electron correlation is divided into static and dynamic correlations; 
\begin{description}
    \item[Static correlation:] contribution from bond-dissociation, excited state, or near degeneracy of electronic configurations such as a singlet diradical ${\rm CH_2}$.
    \item[Dynamical correlation:] contribution from the Coulomb repulsion.
\end{description}
The static correlation can be treated as the multiple Slater determinants such as the multi-configurational self-consistent field (MCSCF), which indicates the elimination of the HF approximation. The dynamic correlation functions as the method to eliminate the effect of the mean-field approximation. Based on our observations, the static and dynamical correlations are not clearly distinct. 

\subsection{Second quantized approach to quantum chemistry}
Let us consider the second quantization form of the electronic Hamiltonian (\ref{eHam}) for the basis of the MO $\{ \ket{\tilde{\varphi}_{i}} \}$ solved by the HFR equation (\ref{HFRM}) as 
\begin{equation}
    \ket{\tilde{\varphi}_{i}} = \sum_{k=1}^{M} \tilde{c}_{ki} \ket{\chi_k},
\end{equation}
where the coefficient $\tilde{c}_{ki}$ is obtained by Eq. (\ref{HFRM}). The number of MOs $\{ \ket{\tilde{\varphi}_{i}} \}$ is same as that of the AO, $M$, which is more than one of the electrons $N_{e}$. Moreover, it is easy to compare $\bkt{\tilde{\varphi}_{i}}{\tilde{\varphi}_{j}} = \delta_{ij}$ to the Kronecker delta $\delta_{ij}$. Then, the MO $\{ \ket{\tilde{\varphi}_{i}} \}$ can be regarded as the complete orthogonal basis of the approximated Fock space to represent quantum many-body system. For the $\ket{\tilde{\varphi}_{i}}$ MO, the fermionic creation and annihilation operators, $\hat{c}^\dagger_i$ and $\hat{c}_i$, satisfy the following equation, 
\begin{equation}
    \ket{\tilde{\varphi}_{i}} = c^\dagger_i \ket{vac}, \
    \hat{c}_i \ket{vac} = 0, \
    [\hat{c}_i, \hat{c}^\dagger_j]_{+}  = \delta_{ij}, \
    [\hat{c}_i, \hat{c}_j]_{+} = 0, \
    [\hat{c}^\dagger_i, \hat{c}^\dagger_j]_{+} = 0
\end{equation}
where $\ket{vac}$ is the vacuum state and $[A, B]_{+} := AB + BA$ is the anti-commutation relationship. Therefore, the electronic Hamiltonian (\ref{eHam}) can be converted to: 
\begin{equation}
    \label{e2Ham}
    \tilde{H}_e = \sum^{M}_{p,q} h_{pq} \hat{c}^\dagger_p \hat{c}_q + \sum^{M}_{p,q,r,s} h_{pqrs} \hat{c}^\dagger_p \hat{c}^\dagger_q \hat{c}_r \hat{c}_s,
\end{equation}
where the one- and two-electron integrals are 
\begin{align}
    h_{pq} & = \int d \vec{x} \tilde{\varphi}^\ast_{p} (\vec{x})
    \left( - \frac{\hbar}{2 m_e} \nabla^2  
    - \frac{1}{4 \pi \epsilon_0} \sum_{I=1}^{N_{nucl}} \frac{Z_I q_e^2}{|\vec{X}_I - \vec{x}|} \right)
    \tilde{\varphi}_{q} (\vec{x}), \\
    h_{pqrs} & = \frac{q_e^2}{4 \pi \epsilon_0} \iint d \vec{\tilde{x}} d \vec{x} \frac{ \tilde{\varphi}^\ast_{p} (\vec{\tilde{x}}) \tilde{\varphi}^\ast_{q} (\vec{x}) \tilde{\varphi}_{r} (\vec{\tilde{x}}) \tilde{\varphi}_{s} (\vec{x})}{|\vec{\tilde{x}} - \vec{x}|},
\end{align}
respectively. This Hamiltonian depends on the basis-set approximation and is slightly different from the original electronic Hamiltonian (\ref{eHam}).

\subsection{Full configuration interactions (full CI)} \label{sec:fullci}
\begin{figure}[t]
    \centering
    \includegraphics{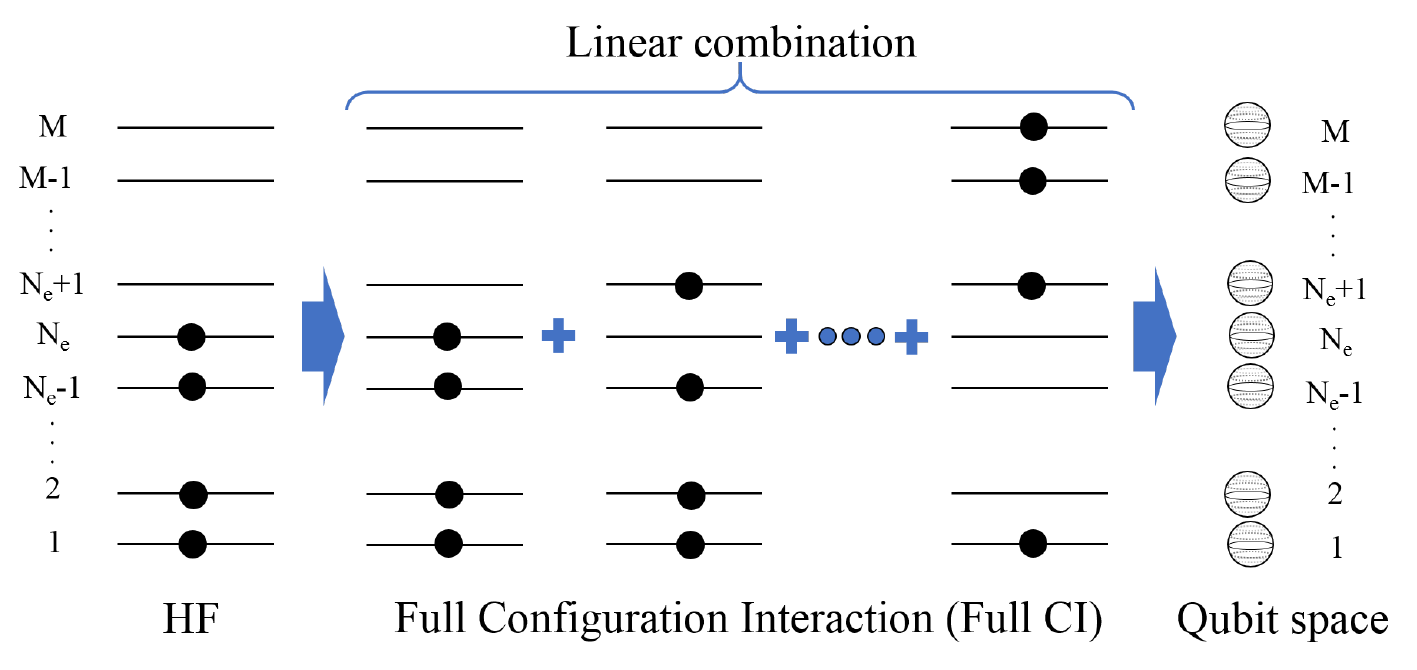}
    \caption{Schematic figure on the molecular orbitals (MOs) on the HF ground state, the full CI, and the mapping to the qubit space.}
    \label{fig:mo}
\end{figure}
Since the $N_e$ electronic state (\ref{slater_rep}) occupies the $N_e$ MOs, the HF ground state, $\ket{\Phi_0}$, is represented by
\begin{equation}
    \ket{\Phi_0} = \hat{c}^\dagger_{N_e} \cdots \hat{c}^\dagger_2 \hat{c}^\dagger_1 \ket{vac}.
\end{equation}
This is depicted in Fig.~\ref{fig:mo}. 

The HF ground state $\ket{\Phi_0}$ is not the ground state of the electronic Hamiltonian (\ref{e2Ham}) due to the electron correlation. To solve this ground state, the {\it correlated} trial states $\ket{\Psi_{CI}}$ without normalization are defined as 
\begin{equation}
    \ket{\Psi_{CI}} = \left( 1 + \sum^{J}_{I = 1} \hat{C}_I \right) \ket{\Phi_0} = \ket{\Phi_0} + \sum^{J}_{I = 1} \frac{1}{(I !)^2} \sum^{N_e}_{i,j,k,\dots} \sum^{N_v}_{a,b,c,\dots} {\mathsf c}^{a,b,c,\dots}_{i,j,k,\dots} \ket{\Phi^{a,b,c,\dots}_{i,j,k,\dots}},
\end{equation}
where the $I$-electron excitation operator is defined as 
\begin{equation}
    \hat{C}_I \equiv \frac{1}{(I !)^2} \sum^{N_e}_{i,j,k,\dots} \sum^{N_v}_{a,b,c,\dots} {\mathsf c}^{a,b,c,\dots}_{i,j,k,\dots} \hat{c}^\dagger_a \hat{c}^\dagger_b \hat{c}^\dagger_c \cdots \hat{c}_k \hat{c}_j \hat{c}_i,
\end{equation}
where the unknown coefficients are $\left\{ {\mathsf  c}^{a,b,c,\dots}_{i,j,k,\dots} \right\}$ and the number of the virtual orbitals are $N_v := M - N_e$. The optimized coefficients are numerically solved by minimizing the trial energy as 
\begin{equation}
    E_{CI} \left(\left\{ {\mathsf c}^{a,b,c,\dots}_{i,j,k,\dots} \right\} \right) = \frac{\bra{\Psi_{CI}} \tilde{H}_e \ket{\Psi_{CI}}}{\bk{\Psi_{CI}}}.
\end{equation}
When all the electron excitation operators are considered, i.e., when $J = N_e$, the solution is termed a full configuration interaction (full CI or FCI). It is denoted as $\ket{\Psi_{FCI}}$. On $J < N_e$, this is also called a truncated CI. When $J=1$ and $J=2$, this is often denoted as CIS and CISD, respectively.

\begin{figure}[ht]
    \centering
    \includegraphics{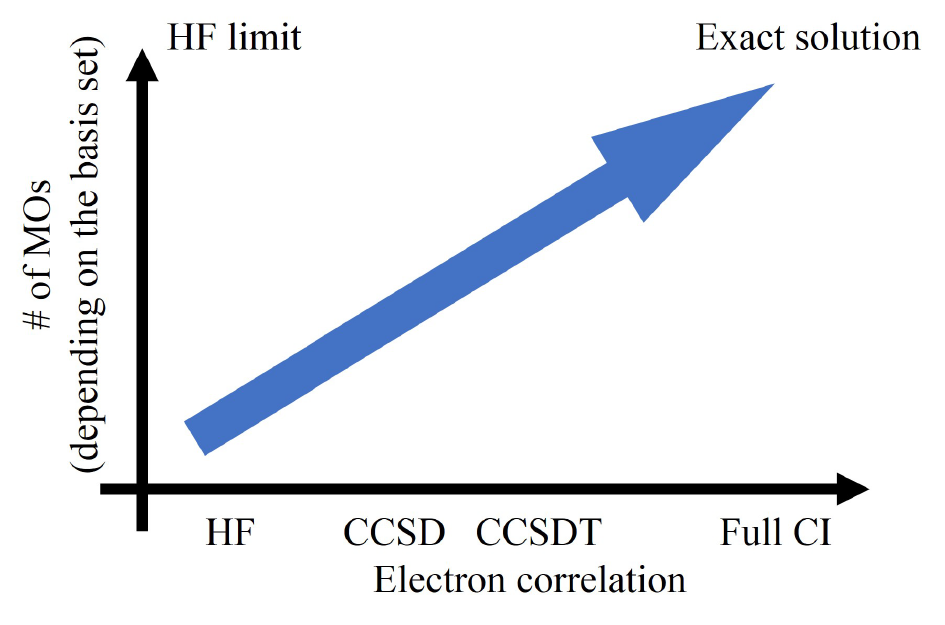}
    \caption{Schematic relationships on the accuracy for the solution of the original Hamiltonian (\ref{eHam}) between the numbers of configuration interaction and those of MOs determined by the basis set.}
    \label{fig:basis_exact}
\end{figure}
As stated before, this approach depends on the basis set approximation. Although the target accuracy of the numerical solution depends on an application as discussed in Sec.~\ref{sec:accuracy}, the full CI is not equivalent to the exact solution of the Hamiltonian (\ref{eHam}), as seen in Fig.~\ref{fig:basis_exact}. For example, the energies with several basis sets are compared for the hydrogen molecule. The ground-state energies with STO-3G and 6-31G are evidently different. Those full CI solutions are still different from the exact solution of the Hamiltonian (\ref{eHam})~\cite{Kolos}.
\begin{figure}[H]
    \centering
    \includegraphics[width=10cm]{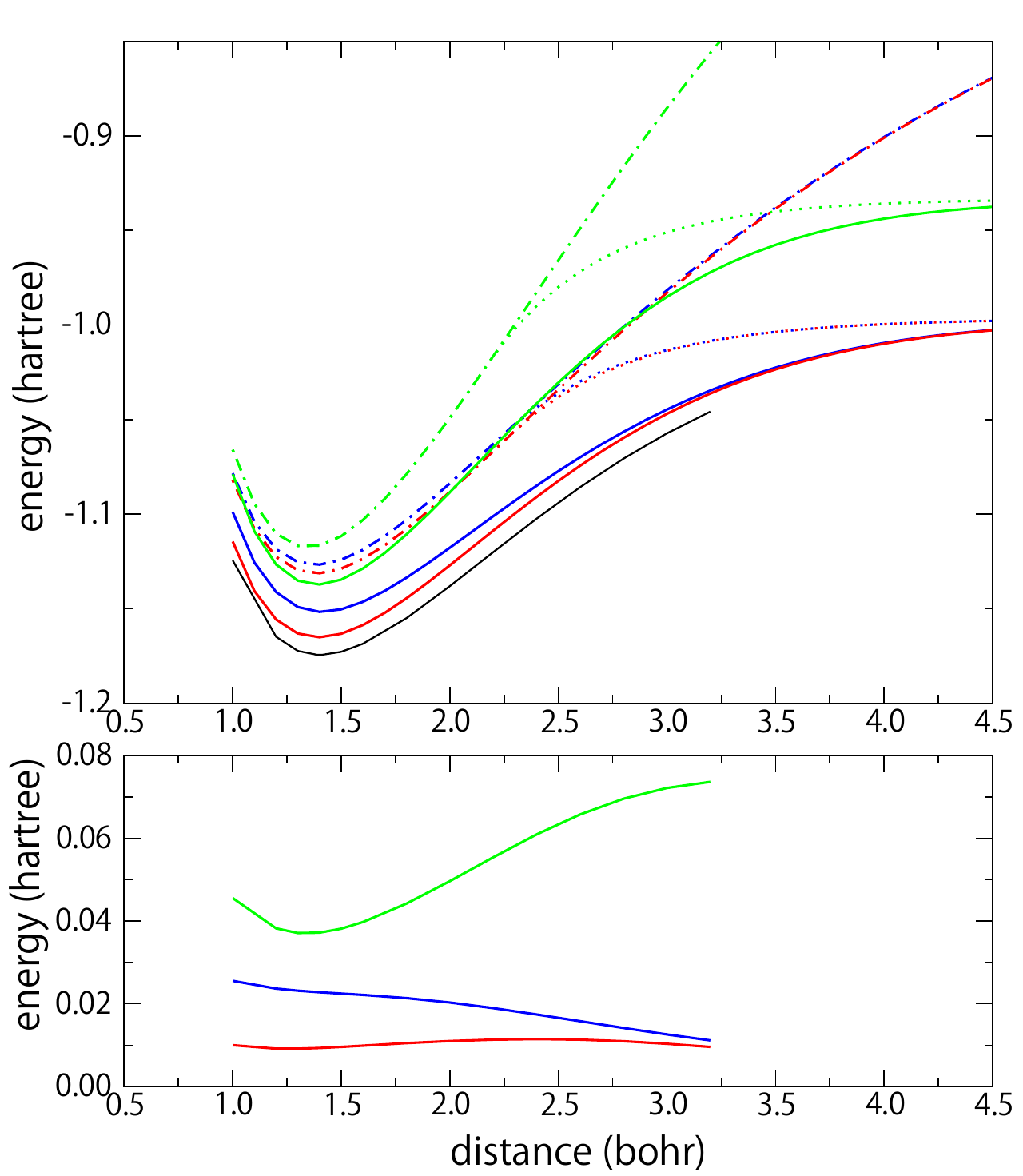}
    \caption{(top) Potential energy curves for the ${\rm H_2}$ of the RHF (broken-line dots), UHF (dots), full CI (line), and with the basis sets, STO-3G (green), 6-31G (blue), and 6-31G**(red), and the exact solution (black)~\cite{Kolos}. (bottom) The difference between the full CI of several basis sets and the exact result~\cite{Kolos}. It is noted that the data points are interpolated.}
    \label{fig:basis}
\end{figure}

\subsection{Coupled-Cluster (CC) theory} \label{sec:cc}
As a different form of the {\it correlated} trial function, a cluster operator $\hat{T}$ is defined as
\begin{equation}
    \ket{\Psi_{CC}} = e^{\hat{T}} \ket{\Phi_0} = e^{\sum^{J}_{I = 1} \hat{T}_I} \ket{\Phi_0},
\end{equation}
where  
\begin{equation}
    \hat{T}_I \equiv \frac{1}{(I !)^2} \sum^{N_e}_{i,j,k,\dots} \sum^{N_v}_{a,b,c,\dots} {\mathsf t}^{a,b,c,\dots}_{i,j,k,\dots} \hat{c}^\dagger_a \hat{c}^\dagger_b \hat{c}^\dagger_c \cdots \hat{c}_k \hat{c}_j \hat{c}_i,
\end{equation}
where the unknown cluster amplitudes are $\left\{ {\mathsf  t}^{a,b,c,\dots}_{i,j,k,\dots} \right\}$. The relationship to the configuration interaction is 
\begin{align}
    \hat{C}_1 & = \hat{T}_1, \\
    \hat{C}_2 & = \hat{T}_2 + \frac{1}{2} \left( \hat{T}_1 \right)^2, \\
    \hat{C}_3 & = \hat{T}_3 + \hat{T}_1 \hat{T}_2 + \frac{1}{6} \left( \hat{T}_1 \right)^3.
\end{align}
In principle, this representation can symbolize the full CI. 

The formal solution for the eigenvalue and the eigenstate problem of the Hamiltonian (\ref{e2Ham}) is expressed as 
\begin{equation}
    \tilde{H}_e \ket{\Psi_{CC,0}} = \tilde{H}_e e^{\hat{T}} \ket{\Phi_0} = E_{CC} e^{\hat{T}} \ket{\Phi_0}. 
\end{equation}
We can then obtain the coupled-cluster equation as 
\begin{align}
    \bra{\Phi_0} e^{- \hat{T}} \tilde{H}_e e^{\hat{T}} \ket{\Phi_0} & = E_{CC}, \label{cceq} \\
    \bra{\Phi^{a,b,c,\dots}_{i,j,k,\dots}} e^{- \hat{T}} \tilde{H}_e e^{\hat{T}} \ket{\Phi_0} & = 0, \label{cceq2}
\end{align}
where $e^{- \hat{T}} e^{\hat{T}} = 1$ is used. It is noted that the orthogonality condition $\bkt{\Phi^{a,b,c,\dots}_{i,j,k,\dots}}{\Phi_0} = 0$ is also used. Further, we obtain the following terminating expansion as  
\begin{align}
    e^{- \hat{T}} \tilde{H}_e e^{\hat{T}} = \tilde{H}_e & + [\tilde{H}_e, \hat{T}] + \frac{1}{2!} [[\tilde{H}_e, \hat{T}], \hat{T}] \notag \\ & + \frac{1}{3!} [[[\tilde{H}_e, \hat{T}], \hat{T}], \hat{T}] + \frac{1}{4!} [[[[\tilde{H}_e, \hat{T}], \hat{T}], \hat{T}], \hat{T}] 
\end{align}
with the commutation relationship $[A,B] = AB - BA$, which is termed a linked diagram theorem. Therefore, Eqs.~(\ref{cceq}, \ref{cceq2}) can be reduced to the simultaneous equations. As a variant of the coupled-cluster method, the variational coupled-cluster (VCC) method was proposed to variationally minimize the trial energy, $E_{VCC}$, defined as 
\begin{equation}
    E_{VCC} = \frac{\bra{\Phi_0} e^{\hat{T}^\dagger} \tilde{H}_e e^{\hat{T}} \ket{\Phi_0}}{\bra{\Phi_0} e^{\hat{T}^\dagger} e^{\hat{T}} \ket{\Phi_0}}.
\end{equation}
In addition, the unitary coupled-cluster (UCC) was similarly proposed to variationally minimize the trial energy $E_{UCC}$ is described by 
as\begin{equation}
    E_{UCC} = \bra{\Phi_0} e^{\hat{T}^\dagger - \hat{T}} \tilde{H}_e e^{\hat{T} - \hat{T}^\dagger} \ket{\Phi_0}.
\end{equation}
The unitary operator can be directly implemented on the quantum computer. The UCC approaches are often used. In principle, these approaches satisfy the full CI but require the non-terminating expansion due to the Baker-Hausdorff-Campbell formula. In the case of the truncated coupled cluster state, a difference occurs on the computational accuracy, as reported in Refs.~\cite{Cooper,Shiozaki,Hodecker}. Compared to the truncated configuration interaction, a {\it size consistency}, which means that a quantum state represented in the two divided subsystems should be the same as one in the whole system, is satisfied. There are several well-known correction methods on this size consistency, which are detailed in Ref.~\cite{Hirata}.
It is observed that UCCSD, $\hat{T} = \hat{T}_1 + \hat{T}_2$, is often used as the quantum computational algorithm that will be discussed later but its computational accuracy is different from that of CCSD. The number of cluster amplitudes is ${\mathcal O} (N^2_e N^2_v)$. Further technical discussions are provided in Ref.~\cite{Helgaker}.

\section{Classical Preparation for Quantum Computing Algorithm} \label{sec:qubit}
A basic structure of a quantum computer~\cite{DiVincenzo} consists of an integrated two-level quantum system, which is called a qubit; this system comprises a long relevant decoherent time, a universal quantum gate that is composed of the single- and double-qubit gates, and a qubit-specific measurement. Furthermore, initialized qubits are well prepared. Therefore, Eq. (\ref{e2Ham}) is not directly implemented in the quantum computer. We need the MO to be within the framework of the basis set approximation of the $M$-qubit system. This method is called a fermion-to-qubit mapping or qubit mapping. After a fermion-to-qubit mapping, the Hamiltonian is formally described by 
\begin{equation}
    {\mathcal H}_e = \sum_{i_1, i_2, \cdots, i_M} \alpha_{i_1, i_2, \cdots, i_M} \hat{\sigma}_{i_1} \otimes \hat{\sigma}_{i_2} \otimes \dots \otimes \hat{\sigma}_{i_M},
    \label{Pauli}
\end{equation}
where $i_1, i_2, \cdots, i_M \in \{ 0, 1=x, 2=y, 3=z \}$ with $\hat{\sigma}_0 \equiv I$. It is noted that $\hat{\sigma}_{i_1} \otimes \hat{\sigma}_{i_2} \otimes \dots \otimes \hat{\sigma}_{i_M}$ is often called a Pauli-operator string. 

A fermion-to-qubit mapping is a one-to-one basis change from the ferminionic basis to qubit described by 
\begin{equation}
    \ket{f_{M-1}, f_{M-2}, \cdots, f_0} \to \ket{q_{M-1}, q_{M-2}, \cdots, q_0}
\end{equation}
In the occupation-number preserving case, this is called a Jordan-Wigner (JW) transformation~\cite{JW} described by 
\begin{equation}
    q_k = f_k \in \{ 0, 1 \}.
\end{equation}
On acting the fermionic operator, $\hat{c}^\dagger_j$ or $\hat{c}_j$, to a MO $\ket{f_{M-1}, f_{M-2}, \cdots, f_j, f_{j-1}, \cdots, f_0}$ with $f_k \in \{0, 1 \}$ in the second quantized form, we obtain 
\begin{align}
    \hat{c}^\dagger_j \ket{f_{M-1}, \cdots, 1, f_{j-1}, \cdots, f_0} & = 0, \label{qubitization1} \\
    \hat{c}^\dagger_j \ket{f_{M-1}, \cdots, 0, f_{j-1}, \cdots, f_0} & = (-1)^{\sum^{j-1}_{k=0} f_k} \ket{f_{M-1}, \cdots, 1, f_{j-1}, \cdots, f_0} \\
    \hat{c}_j \ket{f_{M-1}, \cdots, 1, f_{j-1}, \cdots, f_0} & = (-1)^{\sum^{j-1}_{k=0} f_k} \ket{f_{M-1}, \cdots, 0, f_{j-1}, \cdots, f_0}, \\
    \hat{c}_j \ket{f_{M-1}, \cdots, 0, f_{j-1}, \cdots, f_0} & = 0. \label{qubitization2}
\end{align}
This fact is to delocalize the parity information. On undergoing qubitization, the fermionic operator, $\hat{c}^\dagger_j$ or $\hat{c}_j$, should also be converted to satisfy the properties (\ref{qubitization1})--(\ref{qubitization2}). In the case of JW transformation, the fermionic operator, $\hat{c}^\dagger_j$ or $\hat{c}_j$, 
\begin{align}
    \hat{c}^+_j & = I \otimes I \otimes \cdots \otimes \hat{Q}^+_j \otimes \hat{\sigma}_z \otimes  \cdots \otimes \hat{\sigma}_z, \\
    \hat{c}^-_j & = I \otimes I \otimes \cdots \otimes \hat{Q}^-_j \otimes \hat{\sigma}_z \otimes \cdots \otimes \hat{\sigma}_z,
\end{align}
where $\hat{Q}^+_j \equiv \kbt{1}{0} = \frac{1}{2}(\hat{\sigma}_{x,j} - i \hat{\sigma}_{y,j})$ and $\hat{Q}^-_j \equiv \kbt{0}{1} = \frac{1}{2}(\hat{\sigma}_{x,j} + i \hat{\sigma}_{y,j})$. After this operator transformation, the systematic calculation of Eq.~(\ref{Pauli}) can be executed. It is remarked that the number of Pauli operators is less than 4; this value does not include the identity operator in each Pauli string of Eq.~(\ref{Pauli}) transformed from Eq. (\ref{e2Ham}).

For other fermion-to-qubit mapping methods, in the parity preserving case, this is called a parity encoding~\cite{parity}, which is described by
\begin{equation}
    q_k = \left[ \sum^k_{i=0} f_i \right] \ ({\rm mod} \ 2).
\end{equation}
In the hybridization case between the occupation number and parity information, this is called a Bravyi-Kitaev (BK) transformation~\cite{BK} described by 
\begin{equation}
    q_k = \left[ \sum^k_{i=0} \beta_{ki} f_i \right] \ ({\rm mod} \ 2),
\end{equation}
where the BK matrix $[ \beta_{ki} ]$ is recursively defined by 
\begin{align}
    \beta_1 & = [1], \\
    \beta_{2^j} & = \left[ \begin{array}{cc} \beta_{2^j} & 0 \\ A & \beta_{2^j} \end{array} \right],
\end{align}
where the $2^j$-order square matrix $A$ is defined by 
\begin{equation}
    A = \left[ \begin{array}{ccc} 0 & \dots & 0 \\ \vdots & \ddots & \vdots \\ 0 & \dots & 0 \\ 1 & \dots & 1 \end{array} \right].
\end{equation}
The JW transformation localizes the occupation number for MO, but not the parity information. In contrast, the parity transformation is localizes the parity information, but not the occupation number for MO. The BK transformation partially localizes the occupation number for MO and parity information. From the gate-count viewpoint, the fermion-to-qubit mapping methods are compared in Ref.~\cite{Tranter}. Further generalization of the BK matrix can be considered. Inspired by the data structure and graph theory, several theoretical studies are still developing~\cite{BK1,BK2}. This should be connected to the quantum-computer compiler design~\cite{Chong} to implement this to a real-hardware device.
\section{Quantum Computing Algorithm in Quantum Device} \label{sec:quantum}
\begin{figure}[t]
    \centering
    \includegraphics{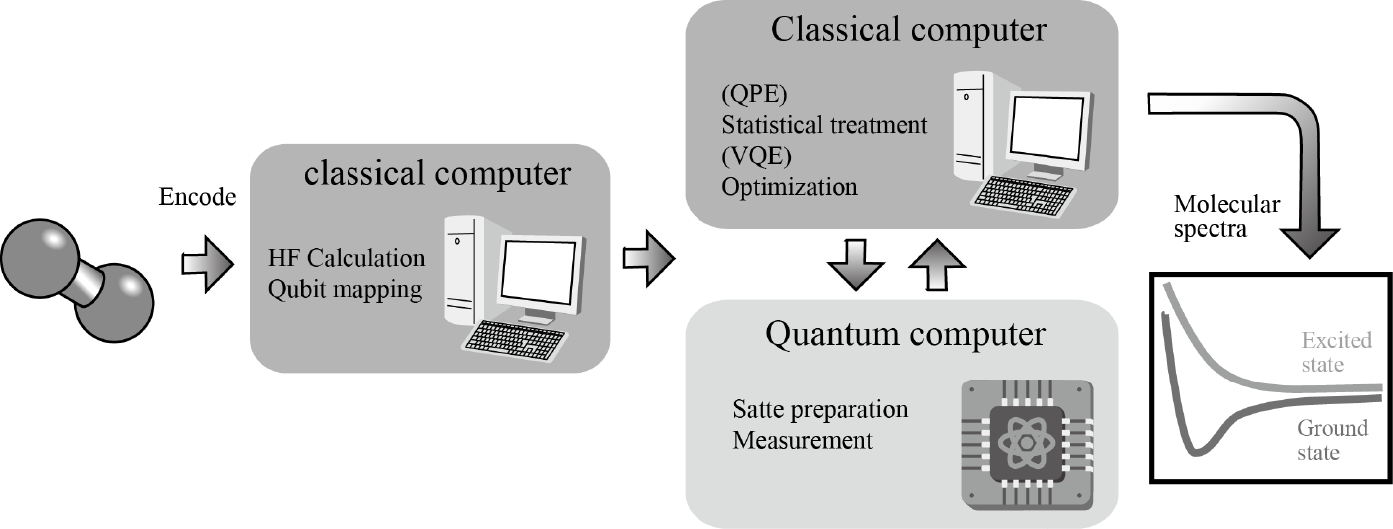}
    \caption{Schematic flow of quantum computing algorithm for quantum chemistry calculation.}
    \label{fig:sqc}
\end{figure}

In the previous sections, we used the classical treatment to accomplish the post-HF method by quantum computers. Solving the eigenvalues and these eigenstates of the qubit Hamiltonian (\ref{Pauli}) with the given coefficients from the fermion-to-qubit mapping is a QMA-complete problem in quantum computational complexity since this Hamiltonian is 4-local Hamiltonian~\cite{Kempe}. The complexity class QMA, Quantum Merlin Arthur, is the quantum analog of the nonprobabilistic complexity class NP, nondeterministic polynomial time, which is a set of decision problems whose answers are verifiable by the deterministic Turing machine in polynomial time. QMA is contained in PP, which is the class of decision problems solvable by a probabilistic Turing machine in polynomial time, but it includes NP. Furthermore, QMA-complete means that any problems in QMA are transformed to the QMA-complete problem by the deterministic Turing machine in polynomial time. Even quantum algorithms do not perfectly solve this eigenvalues and these eigenstates of Eq.~(\ref{Pauli}) in polynomial time with respect to the number of basis sets. Therefore, quantum algorithms in quantum computational chemistry often use heuristic or probabilistic methods. Our schematic treatment is depicted in Fig.~\ref{fig:sqc}. We will subsequently explain two well-known quantum algorithms: quantum phase estimation and variational quantum eigensolver. 

\subsection{Quantum phase estimation (QPE)} \label{sec:qpe}
Let us consider the general description of the time evolution by the Hamiltonian ${\mathcal H}_e$ under a given initial state $\ket{\psi (0)}$ as
\begin{align}
    \ket{\psi (t)} & = e^{-i \frac{{\mathcal H}_e}{\hbar} t} \ket{\psi (0)} \equiv U_t \ket{\psi (0)} \\ & = \sum_{j=0}^{M-1} e^{- i \frac{\epsilon_j}{\hbar} t} \bkt{\epsilon_j}{\psi (0)} \ket{\epsilon_j} = \sum_{j=0}^{M-1} a_j e^{- i (2 \pi) \phi_j} \ket{\epsilon_j} 
\end{align}
with ${\mathcal H}_e = \sum_{j=0}^{M-1} \epsilon_j \kb{\epsilon_j}$ and $a_j \equiv \bkt{\epsilon_j}{\psi (0)}$. The phase $\phi_j \equiv \epsilon_j t / (2 \pi \hbar) \in [ 0, 1)$ has the information on the energy eigenstate of the FCI Hamiltonian. Therefore, a method to estimate the phase $\phi_j$ using quantum algorithms is called a quantum phase estimation. Since the output of quantum computers for each qubit uses the binary value, the $\phi_j$ phase is expanded as the binary value.
\begin{equation}
    \phi_j = \sum_{i = 1}^{\tilde{N} = \infty} \frac{\phi^{(2)}_i}{2^i} = 0. \phi^{(2)}_{1,j} \phi^{(2)}_{2,j} \cdots \phi^{(2)}_{\tilde{N}, j} \cdots.
\end{equation}
Since the quantum computer has a finite qubit resource, we have to consider the truncation of the binary phase value; $\tilde{N}$ is fixed as the finite value, which corresponds to the round-off error. The phase periodicity over $2 \pi$ shows the multiplication of the same time evolution as 
\begin{equation}
    U^{2^k}_t \ket{\psi (0)} = \sum_{j=0}^{M-1} a_j e^{- i (2 \pi) 2^k \phi_j} \ket{\epsilon_j},
\end{equation}
The estimated phase is converted to 
\begin{align}
    2^k \phi_j & = \phi^{(2)}_{1,j} \cdots \phi^{(2)}_{k,j}.\phi^{(2)}_{(k+1),j} \phi^{(2)}_{(k+2),j}  \cdots \phi^{(2)}_{\tilde{N},j} \cdots \notag \\ 
    & \simeq 0 .\phi^{(2)}_{(k+1),j} \phi^{(2)}_{(k+2),j}  \cdots \phi^{(2)}_{\tilde{N},j} \cdots.
\end{align}
Here, the last equality indicates the equivalence in the terms of the phase estimation.

In the basic scheme of the QPE for quantum chemistry calculation, the $\tilde{N}$ ancilla qubits are initially prepared to the quantum state $\ket{+}_{1} \cdots \ket{+}_{\tilde{N}}$ with $\ket{+} = \frac{1}{\sqrt{2}} (\ket{0} + \ket{1})$. The following notation is introduced:  
\begin{equation}
    \frac{1}{\sqrt{2^{\tilde{N}}}} \sum_{x} \ket{x} \equiv \ket{+}_1 \cdots \ket{+}_k \cdots \ket{+}_{\tilde{N}}, \ \ 
    \ket{\psi(0)} = \sum^{M-1}_{j=0} a_j \ket{\epsilon_j}.
\end{equation}
The controlled--$U^{2^k}_t$ operations between the $k$-th ancilla qubit and a represented state $\ket{\psi}$ in the reverse order from $\tilde{N}$-th ancilla qubit to the first ancilla one is sequentially operated to obtain 
\begin{align}
    \ket{+}_1 \cdots \ket{+}_k \cdots \ket{+}_{\tilde{N}} \ket{\psi (0)} 
    & \to \frac{1}{\sqrt{2^{\tilde{N}}}} 
    \sum_{x} \sum^{M-1}_{j=0} a_j e^{-i 2 \pi \phi_j x} \ket{x} \ket{\epsilon_j}  \notag \\
    & \xrightarrow{{\mathrm{QFT}}^{-1}} \sum_{j} a_j \ket{f_2(\phi_j)} \ket{\epsilon_j},
\end{align}
where $f_2(\phi_j)$ is the binary representation of the eigenvalue $\phi_j$. ${\mathrm{QFT}}^{-1}$ is the inverse Fourier transformation acting on the ancilla qubits. Finally, the measurement to the $\tilde{N}$ ancilla qubits is taken to obtain the desired eigenvalue $f_2(\phi_j)$ with the probability $|a_j|^2$. This procedure can be repeatedly taken to increase the success probability to obtain the desired eigenvalue. The weight of $|a_j|^2$ depending on the choice of the initial state should be high, but should not be the perfect one. This algorithm is probabilistic. 

In addition to the hardware imperfection, this algorithm theoretically contains the following errors: (i) algorithmic error and (ii) statistical error. In terms of the algorithmic error, the unitary time evolution cannot be flawlessly operated due to the Trotter-Suzuki error. There are several theoretical developments with regard to the error analysis of the higher order operations~\cite{Campbell,Childs}. As another methodology, qubitization was recently proposed, which is inspired by the Grover search~\cite{Low}. There are several treatments to reduce this algorithm error depending on the basis set approximation as seen in Table~\ref{table:qpe_cost}. There is a trade-off relationship between this error and the computational time.
\begin{table}[t]
    \centering
    \begin{tabular}{c|c|c|c}
        \hline
         Ref. & Basis Set & Method &  Total T-Count \\ \hline
         \cite{Wecker} & Gaussians & Trotterization & $\mathcal{O} \left( {\tilde{N}}^{10} \log \left( 1/ \epsilon \right) / \epsilon^{3/2} \right)$ \\ 
         \cite{Low} & Gaussians & Qubitization & $\tilde{\mathcal{O}} \left( {\tilde{N}}^5 / \epsilon \right)$ \\
         \cite{Babbush} & Plane-wave &  Qubitization & $\mathcal{O} \left( \left( {\tilde{N}}^{3} + {\tilde{N}}^2 \log \left( 1/ \epsilon \right) \right) / \epsilon \right)$ \\ \hline
    \end{tabular}
    \caption{The lowest T complexity QPE algorithms with the algorithmic error $\epsilon$. It is noted that $\tilde{\mathcal{O}}( \cdot )$ indicates an upper bound ignoring polylogarithmic factors.}
    \label{table:qpe_cost}
\end{table}
In addition, the statistical error indicates that the successful phase estimation should be probabilistic since the single-shot output from quantum computers is probabilistic. Therefore, statistical treatments are required after consistently running the same quantum circuit. This estimation error is based on the initial prepared quantum state $\ket{\psi (0)}$, which is usually set as the HF ground state $\ket{\Phi_0}$. This is because the overlap between the HF ground state and the FCI ground state is high. The precision-guaranteed QPE algorithm is proposed with the help of the hypothetical test~\cite{Shikano}. Furthermore, to reduce the number of the ancilla qubits, the Kitaev QPE~\cite{Kitaev} and the iterative QPE~\cite{Dobsicek} algorithms that facilitate the use of one ancilla qubit are developing~\cite{Svore,Wiebe,Brien,Berg}.

This QPE algorithm only solves the phase corresponding to the eigenvalues of the FCI Hamiltonian. Under the obtained eigenvalues, we should additionally calculate the corresponding eigenstates. This computational cost is roughly evaluated as $\mathcal{O} (\mathrm{poly} (\mathrm{log} M))$~\cite{classicalalgorithm}.

\subsection{Variational quantum eigensolver (VQE)} \label{sec:vqe}
Let us prepare the parametrized quantum circuit $U(\theta)$, whose construction is discussed later, to obtain the the parametrized quantum state $\ket{\psi (\vec{\theta}_k)}$ from the initial quantum state $\otimes_{m=0}^{M-1} \ket{0}_m$.
We obtain the trial energy $E_{\vec{\theta}_k}$ as 
\begin{equation}
    E_{\vec{\theta}_k} = \bra{\psi (\vec{\theta}_k)} {\mathcal H}_{e} \ket{\psi (\vec{\theta}_k)}.
\end{equation}
This trial energy should be minimized by a variational method to update the parametrized quantum state $\ket{\psi (\vec{\theta}_k)}$. For the rough convergence of the trial energy, $E_{\vec{\theta}_k} \simeq E_{con}$, the ground state and its energy might be calculated. The aforementioned method is called a variational quantum eigensolver (VQE)~\cite{Peruzzo,Yung}. This schematic procedure is depicted in Algorithm~\ref{algo} based on the theoretical basis~\cite{VQEupdate}. In the line \ref{classicalevalution}, there is a chance to set the classical estimation procedure. In the line \ref{updating}, there is also a chance to choose the parameters' updating method, which can be taken as the mathematical optimization problem~\cite{Nocedal}, such as gradient decent and stochastic optimization. There is still an open problem for finding the systematic strategy on VQE.
\begin{algorithm}[t] 
\caption{Minimizing $E_{\vec{\theta}_k} = \bra{\psi (\vec{\theta}_k)} {\mathcal H}_{e} \ket{\psi (\vec{\theta}_k)}$.}
\label{algo}
\begin{algorithmic}[1]
\REQUIRE Parameterized quantum circuits $U(\vec{\theta}_k)$ associated with the parameters $\vec{\theta}_k$
\REQUIRE Initial parameter set $\vec{\theta}_1$
\REQUIRE Updating condition
\REQUIRE Convergence condition
\REQUIRE Maximum iteration step $K_{max}$
\STATE $k \leftarrow 1$
\WHILE{$k < K_{max}$}
\STATE Executing the parameterized quantum circuits associated with the parameters $\vec{\theta}_k$ to obtain the parametrized quantum state $\ket{\psi (\vec{\theta}_k)} = U(\vec{\theta}_k) (\otimes_{m=0}^{M-1} \ket{0}_m)$.
\STATE Evaluating the trial energy $E_{\theta_k} = \bra{\psi (\vec{\theta}_k)} {\mathcal H}_{e} \ket{\psi (\vec{\theta}_k)}$ from the measurement result of quantum-circuit execution. \label{classicalevalution}
\IF{The updating condition is satisfied.} 
\STATE $E_{con} \leftarrow E_{\theta_k}$
\STATE $\ket{\psi (\vec{\theta}_{con})} \leftarrow \ket{\psi (\vec{\theta}_k)}$
\ENDIF
\IF{$\mathrm{min} E_{\theta}$ satisfies the convergence condition.}
\STATE $k \leftarrow K_{max}$
\ELSE
\STATE $k \leftarrow k + 1$
\STATE Updating the trial wavefuction $\ket{\psi (\vec{\theta}_k)}$ by updating the parameter $\vec{\theta}_k$. \label{updating}
\ENDIF
\ENDWHILE
\STATE Obtaining the energy $E_{con}$ and its associated wavefunction $\ket{\psi (\vec{\theta_{con}})}$.
\end{algorithmic}
\end{algorithm}
Since the parametrized quantum state $\ket{\psi (\vec{\theta}_k)}$ represents the $N$ qubit, the number of parameters $\vec{\theta}_k$ requires a $2^N$-dimensional complex vector space to search the entire Hilbert space. 
Therefore, this parameter-update method indicates a combinatorial explosion. Two approaches are often used as the parameter-number restrictions: (i) heuristic approach and (ii) physics-based approach. For the heuristic approach, an initially prepared specific entangled state is often used~\cite{Barkoutsos}, which is commonly considered a hardware-efficient method. This has a drawback in barren plateaus~\cite{McClean}. 
For the physics-based approach, a truncated unitary coupled-cluster (tUCC) method such as UCCSD is often used, as explained in Sec.~\ref{sec:cc}. As previously mentioned, the obtained value cannot theoretically approach the FCI energy, even when the mathematical optimization method is accurately performed. Although there are many combinations of truncation (e.g., tUCC) and iterative methods for mathematical optimization, it is challenging to ensure an optimized solution to estimate the computational speed as mentioned before. This optimization methods are roughly classified into deterministic gradient methods such as gradient decent methods, deterministic Hessian methods such as the Newton method, probabilistic gradient methods such as simultaneous perturbation stochastic approximation (SPSA), heuristic methods such as the Nelder-Mead method, and machine-learning methods. The convergence speed strongly depends on the optimization method~\cite{Nakanishi}. Several theoretical treatments are still under development~\cite{Ryabinkin,Lee,Gard,Matsuzaki,kubler,knolle,sweke}. Furthermore, several algorithms are required to solve the excited-state energies under this framework~\cite{McClean2,Colless,Higgott,Grimsley,Jones,Nakanishi2}. A quantum computational method involving electronic transitions was also proposed~\cite{Parrish}. The equation-of-motion (EOM) approach, which was proposed using the shell model for low-energy nuclear physics~\cite{Rowe}, can compute the energy spectrum of the Hamiltonian combined with the VQE~\cite{Ollitrault}. 

On the other hand, VQE algorithms are expected to be applicable to noisy intermediate scale quantum (NISQ) computers~\cite{Preskill} and to be error resilient. VQE algorithms do not only optimize the energy of the Hamiltonian, but also obtain the high fidelity between the obtained quantum state and the FCI ground state, $F (\ket{\psi (\vec{\theta}_k)}, \ket{\Psi_{FCI}}) := \left| \bkt{\psi (\vec{\theta}_k)}{\Psi_{FCI}} \right|^2$, to obtain the desired quantum state. However, this fidelity cannot be computed because the FCI ground state $\ket{\Psi_{FCI}})$ is uncomputed. To ensure the potential solution by the VQE algorithm, another method is needed. In addition, the error-mitigation methods for the elimination of the hardware error~\cite{Dewes} are often applied to the VQE algorithms to minimize the trial energy~\cite{Li,Temme}. By changing the error rate, the extrapolation technique is applied to virtually eliminate this error. There are several theoretical development~\cite{McArdle,Bonet-Monroig} to be reviewed in Ref.~\cite[Chapter 5]{endo}. By using the error-mitigation method, the final trial quantum state does not correspond to the desired quantum state. Hence, it is necessary to recalculate the quantum state from the obtained energy to be equivalent to the QPE algorithm. 

In the real quantum computational devices, ground-state calculations were executed to be summarized in Ref.~\cite[Table 2]{qc_review} up to 12-qubit calculation~\cite{google_qc}. The excited-state calculation was executed~\cite{Ollitrault}. These benchmark results can be compared with the FCI solutions by the conventional computational technique, and they can be used to evaluate the computational error against the ideal FCI result. Furthermore, the vibrational spectra were also computed in real quantum computational devices~\cite{vib1,vib2,vib3}. As hardware development, a bigger-size quantum chemistry calculation will be computed in the real devices to reach a quantum advantage region.

\section{Conclusion} \label{sec:conclusion}
As an application of quantum computers, the post-HF methods are applied after numerically solving the HF method in conventional computers. The solution of the QPE is given by one of the full CI methods. In the parameterized quantum state, the VQE cannot effectively obtain the full CI solution using polynomial-size parameters for the number of basis sets. In quantum computers, some of the electron correlations are computed. As seen in Fig.~\ref{fig:basis}, there still remains the basis set approximation, even when calculating the full CI solution. During the long history of quantum chemistry, the HF and post-HF methods have been continuously developing as computational methods. Emerging quantum computers are expected to solve the molecular spectra more efficiently. However, even when quantum computers are utilized, several approximations of the HF method remain unchanged. ENIAC, which is the first electronic general-purpose digital computer, pioneered the new computational problems and tools such as the pseudo random number generator and the Monte Carlo method. Hence, the utility of available quantum computers is expected to result in a paradigm shift for computational chemistry like the emergence of the HF method and Kohn--Sham equation of DFT. This will promote an enhanced understanding of the fundamental mechanism or concept of complex molecular dynamics and chemical reactions. 

\section*{Acknowledgement}
The authors thank 
Maho Nakata, 
Takeshi Abe, 
Shumpei Uno, 
Kenji Sugisaki, 
Rudy Raymond,  
and the members of the industry-academia collaborative working team, {\it Quantum Computation for Quantum Chemistry (QC4QC)}, at Quantum Computing Center, Keio University as IBM Q Network Hub at Keio University;
Gao Qi, 
Eriko Watanabe, 
Shoutarou Sudou, 
Takeharu Sekiguchi, 
Eriko Kaminishi, 
Yohichi Suzuki, 
Michihiko Sugawara, 
Shintaro Niimura, 
and Tomoko Ida, 
for their useful suggestions on the manuscript and the discussion. 
Y.S. is grateful to Iwao Ohmine for guiding to molecular science through insightful discussions and Akihito Ishizaki for valuable discussions on physical chemistry.
This work is partially supported by JSPS KAKENHI (Grant Nos. 17K05082, 19K14636, 19H05156, 20H00335, 20H05518, and 20K03885), JST PRESTO (Grant No. JPMJPR17GC) and JST, PRESTO (feasibility study of specific research proposal) (Grant No. JPMJPR19MB). 
K.M.N. thanks IPA for its support through the MITOU Target program. 
H.C.W. is also supported by the MEXT Quantum Leap Flagship Program Grant No. JPMXS0118067285.

\end{document}